\documentclass[sigconf]{acmart}
\AtBeginDocument{%
  }

\usepackage{multirow}
\usepackage{pifont}

\definecolor{cmarkgreen}{HTML}{3C8031}
\definecolor{xmarkred}{HTML}{B6321C}

\newcommand{\cmark}{%
  {\normalfont\raisebox{0.2ex}{%
    \footnotesize\textcolor{cmarkgreen}{\ding{51}}}}%
}

\newcommand{\xmark}{%
  {\normalfont\raisebox{0.2ex}{%
    \footnotesize\textcolor{xmarkred}{\ding{55}}}}%
}

\copyrightyear{2026}
\acmYear{2026}
\setcopyright{cc}
\setcctype{by}
\acmConference[MM '26]{Proceedings of the 34th ACM International Conference on Multimedia}{November 10--14, 2026}{Rio de Janeiro, Brazil}
\acmBooktitle{Proceedings of the 34th ACM International Conference on Multimedia (MM '26), November 10--14, 2026, Rio de Janeiro, Brazil}
\acmDOI{10.1145/3767308.3836081}
\acmISBN{979-8-4007-2213-4/2026/11}

\begin{document}

\title{Geometry-Aware Localized Watermarking for Copyright Protection in Embedding-as-a-Service}

\author{Zhimin Chen}
\orcid{0009-0008-7350-8784}
\email{chenzhm56@mail2.sysu.edu.cn}
\affiliation{
    \institution{School of Computer Science and Engineering, Sun Yat-sen University}
    \city{Guangzhou}
    \state{Guangdong}
    \country{China}
}

\author{Xiaojie Liang}
\orcid{0009-0005-5342-0003}
\email{liangxj53@mail2.sysu.edu.cn}
\affiliation{%
    \institution{School of Computer Science and Engineering, Sun Yat-sen University}
    \city{Guangzhou}
    \state{Guangdong}
    \country{China}
}

\author{Wenbo Xu}
\orcid{0009-0007-8702-4521}
\email{xuwb25@mail2.sysu.edu.cn}
\affiliation{%
    \institution{School of Computer Science and Engineering, Sun Yat-sen University}
    \city{Guangzhou}
    \state{Guangdong}
    \country{China}
}

\author{Yuxuan Liu}
\orcid{0009-0000-7758-9894}
\email{liuyx587@mail2.sysu.edu.cn}
\affiliation{
    \institution{School of Computer Science and Engineering, Sun Yat-sen University}
    \city{Guangzhou}
    \state{Guangdong}
    \country{China}
}

\author{Wei Lu}
\correspondingauthor
\orcid{0000-0002-4068-1766}
\email{luwei3@mail.sysu.edu.cn}
\affiliation{
    \institution{School of Computer Science and Engineering, Sun Yat-sen University}
    \city{Guangzhou}
    \state{Guangdong}
    \country{China}
}

\renewcommand{\shortauthors}{Zhimin Chen, Xiaojie Liang, Wenbo Xu, Yuxuan Liu, and Wei Lu}

\begin{abstract}
  Embedding-as-a-Service (EaaS) has become an important semantic infrastructure for natural language and multimedia applications, but it is highly vulnerable to model stealing and copyright infringement. Existing EaaS watermarking methods face a fundamental robustness--utility--verifiability tension: trigger-based methods are fragile to paraphrasing, transformation-based methods are sensitive to dimensional perturbation, and region-based methods may incur false positives due to coincidental geometric affinity. 
  To address this problem, we propose GeoMark, a geometry-aware localized watermarking framework for EaaS copyright protection. GeoMark uses a natural in-manifold embedding as a shared watermark target, constructs geometry-separated anchors with explicit target--anchor margins, and activates watermark injection only within adaptive local neighborhoods. This design decouples where watermarking is triggered from what ownership is attributed to, achieving localized triggering and centralized attribution. 
  Experiments on four benchmark datasets show that GeoMark preserves downstream utility and geometric fidelity while maintaining robust copyright verification under paraphrasing, dimensional perturbation, and CSE (Clustering, Selection, Elimination) attacks, with improved verification stability and low false-positive risk.
\end{abstract}

\begin{CCSXML}
<ccs2012>
   <concept>
       <concept_id>10002978.10002991.10002996</concept_id>
       <concept_desc>Security and privacy~Digital rights management</concept_desc>
       <concept_significance>500</concept_significance>
       </concept>
   <concept>
       <concept_id>10010147.10010178</concept_id>
       <concept_desc>Computing methodologies~Artificial intelligence</concept_desc>
       <concept_significance>500</concept_significance>
       </concept>
 </ccs2012>
\end{CCSXML}

\ccsdesc[500]{Security and privacy~Digital rights management}
\ccsdesc[500]{Computing methodologies~Artificial intelligence}

\keywords{Model watermarking, model stealing attack, embedding-as-a-service, copyright protection, AI security}

\maketitle

\section{Introduction}

  Large language models (LLMs)~\cite{achiam2023gpt, team2023gemini, yang2025qwen3, touvron2023llama} have demonstrated remarkable capabilities in producing high-quality semantic representations. To avoid the substantial costs of local deployment, developers increasingly deliver these capabilities through Embedding-as-a-Service (EaaS) APIs, such as OpenAI's \texttt{text-embedding-3}\footnote{\url{https://openai.com/index/new-embedding-models-and-api-updates/}}. Beyond traditional natural language processing~\cite{leenv, li2024bellm, fu2025token, cheng2025contrastive, zhao2025tiny, zhang2025uora}, text-based EaaS services have evolved into essential semantic infrastructure for modern multimedia systems. For instance, Multimodal RAG (M-RAG) pipelines heavily rely on text EaaS as semantic anchors to index and retrieve diverse multimedia elements, including parsed document chunks, structured tables, and vision-language image summaries~\cite{r2025multimodalragunstructureddataleveraging, hsiao2025megaragmultimodalknowledgegraphbased, lumer2025comparisontextbasedimagebasedretrieval}. Consequently, safeguarding the intellectual property (IP) of EaaS models is crucial for sustaining the high-quality infrastructure that these expanding multimedia applications depend on.

  However, recent studies~\cite{liu2022stolenencoder, peng-etal-2023-copying} have shown that EaaS systems are highly vulnerable to model stealing attacks. By querying the victim API on an unlabeled corpus and collecting the returned embeddings, an adversary can train a functionally similar surrogate model at only a fraction of the original computational cost. Such attacks allow malicious actors to deploy competing services without authorization, thereby causing substantial economic losses and infringing upon the IP of model providers. These risks make robust copyright protection for EaaS models increasingly important.

\begin{figure}[t]
  \centering
  \includegraphics[width=0.98\columnwidth]{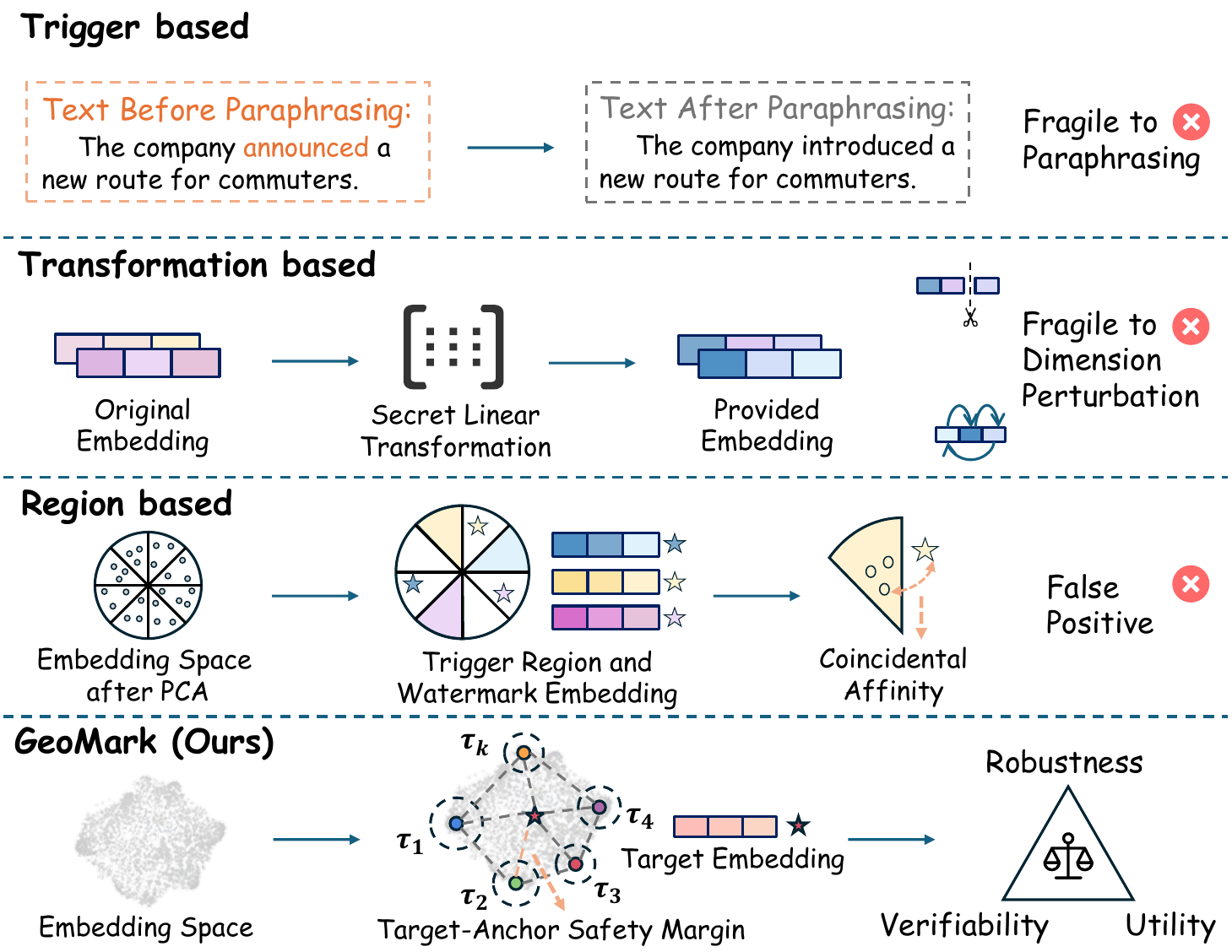}
  \caption{Motivation of GeoMark. Existing EaaS watermarking methods face a robustness--utility--verifiability tension, while GeoMark decouples localized triggering from shared-target attribution to improve robustness and verification reliability with low false-positive risk.}
  \Description{A motivation overview comparing existing EaaS watermarking paradigms with GeoMark, highlighting the robustness--utility--verifiability tension and the proposed distributed-triggering, centralized-attribution design.}
  \label{fig:motivation}
  \end{figure}

  Existing EaaS watermarking defenses face a practical robustness--utility--verifiability tension, as illustrated in Figure~\ref{fig:motivation}, and can be broadly categorized into three paradigms. First, \textit{trigger-based watermarking}~\cite{peng-etal-2023-copying, shetty-etal-2024-warden, wang-etal-2025-robust} relies on fragile surface-form cues, making it vulnerable to paraphrasing~\cite{krishna2023paraphrasing, shetty-etal-2025-wet}. Second, \textit{transformation-based methods}~\cite{shetty-etal-2025-wet} encode ownership via secret transformations, but depend heavily on coordinate structures, rendering them fragile under dimensionality perturbations. Finally, \textit{semantic region-based watermarking}~\cite{yang2025regionmarkerregiontriggeredsemanticwatermarking} activates watermarks based on semantic subregions to resist semantic-preserving attacks. However, because natural embeddings are non-uniformly distributed, region-dependent assignment inherently introduces distance bias~\cite{li2025essencedefenseadaptivesemanticaware}. This bias causes clean embeddings to coincidentally exhibit geometric affinity to assigned targets, accumulating false-positive risks during verification. Simply reverting to a naive shared-target variant is non-trivial, as it often becomes highly vulnerable to the CSE attack. These observations indicate that robust triggering and reliable verification must be designed jointly rather than optimized in isolation.

  To address this challenge, we ask a fundamental question: \textit{Can an EaaS watermark remain robust against diverse removal attacks while preserving embedding utility and still support reliable copyright verification with low false-positive risk?} We answer this question affirmatively with \textbf{GeoMark}, a geometry-aware localized watermarking framework for Embedding-as-a-Service (EaaS). GeoMark first selects a natural in-manifold embedding from the original embedding space as a shared secret watermark target, and then uses farthest point sampling (FPS) to select well-separated anchors that establish explicit geometric safety margins. Around these anchors, GeoMark further calibrates adaptive local neighborhoods so that watermark activation remains localized. Meanwhile, ownership evidence is always aggregated with respect to the same shared watermark target. In this way, GeoMark decouples \emph{where watermarking is triggered} from \emph{what ownership is attributed to}, achieving distributed triggering and centralized attribution.

  In summary, our main contributions are as follows:
  \begin{itemize}
      \item We identify the robustness--utility--verifiability tension in existing EaaS watermarks and show that geometry-aware design is crucial for reliable copyright verification with low false-positive risk.
      \item We propose GeoMark, a geometry-aware localized watermarking framework that decouples localized triggering from shared-target attribution via FPS, improving verification stability and reducing false-positive risk.
      \item Extensive experiments on four benchmark datasets demonstrate that GeoMark preserves embedding utility and geometric fidelity while providing robust copyright verification against diverse watermark removal attacks.
  \end{itemize}

\section{Related Work}

  \subsection{Model Stealing Attacks}

    Model stealing attacks~\cite{10.5555/3241094.3241142, orekondy2019knockoff, DBLP:conf/iclr/KrishnaTPPI20, wallace-etal-2020-imitation, shen2025medical, beethamdual}, also known as model extraction or imitation attacks, aim to replicate a victim model by exploiting black-box query access. By repeatedly querying the target model and collecting its outputs, an adversary can construct a synthetic input--output dataset and train a surrogate model that approximates the victim's behavior at a substantially lower cost. Recent studies have shown that this threat is particularly relevant to Embedding-as-a-Service (EaaS) systems~\cite{liu2022stolenencoder, peng-etal-2023-copying}, where attackers query an embedding API on an unlabeled corpus and train a surrogate encoder using the returned embeddings. This EaaS-oriented query-based extraction setting constitutes the primary threat model studied in this work.

  \subsection{EaaS Watermarks}

    Compared with conventional DNN watermarking settings~\cite{uchida2017embedding, adi2018turning, zhang2018protecting}, EaaS watermarking defenses can be broadly grouped into three paradigms, reflecting different trade-offs among robustness, utility, and verifiability. The first paradigm, \textit{trigger-based watermarking}, injects watermark signals into inputs containing predefined trigger words, as exemplified by EmbMarker~\cite{peng-etal-2023-copying}, WARDEN~\cite{shetty-etal-2024-warden}, and EspeW~\cite{wang-etal-2025-robust}. These methods can effectively inherit ownership evidence under standard extraction, but because their triggering mechanism relies on surface-form cues rather than semantic neighborhoods, they remain vulnerable to paraphrasing attacks. To improve robustness beyond lexical triggers, the second paradigm explores \textit{transformation-based methods}, such as WET~\cite{shetty-etal-2025-wet}, which embed ownership evidence through a secret global transformation in the representation space. However, because such methods depend on preserving the transformed coordinate structure, dimensionality perturbations such as cyclic shifts or truncation can substantially weaken their effectiveness.

    More recent work has shifted toward a \textit{semantic region-based approach}, which partitions the embedding space into semantic subregions and activates watermarking according to region membership~\cite{yang2025regionmarkerregiontriggeredsemanticwatermarking}. This design improves robustness against semantic-preserving attacks, but also makes verification more delicate. Since natural embeddings are not uniformly distributed in semantic space, region-dependent watermark assignment may introduce coincidental geometric affinity between clean embeddings and assigned targets~\cite{li2025essencedefenseadaptivesemanticaware}. When verification aggregates evidence across multiple regions, such affinity may accumulate and increase false-positive risk. Meanwhile, simply reverting to a shared-target variant is not a trivial solution, since naive single-target designs can become more vulnerable to the CSE attack. These limitations motivate GeoMark, a geometry-aware localized approach to robust and reliable model watermarking for EaaS copyright protection.

\section{Methodology}

  \begin{figure*}[t]
  \centering
  \includegraphics[width=0.96\textwidth]{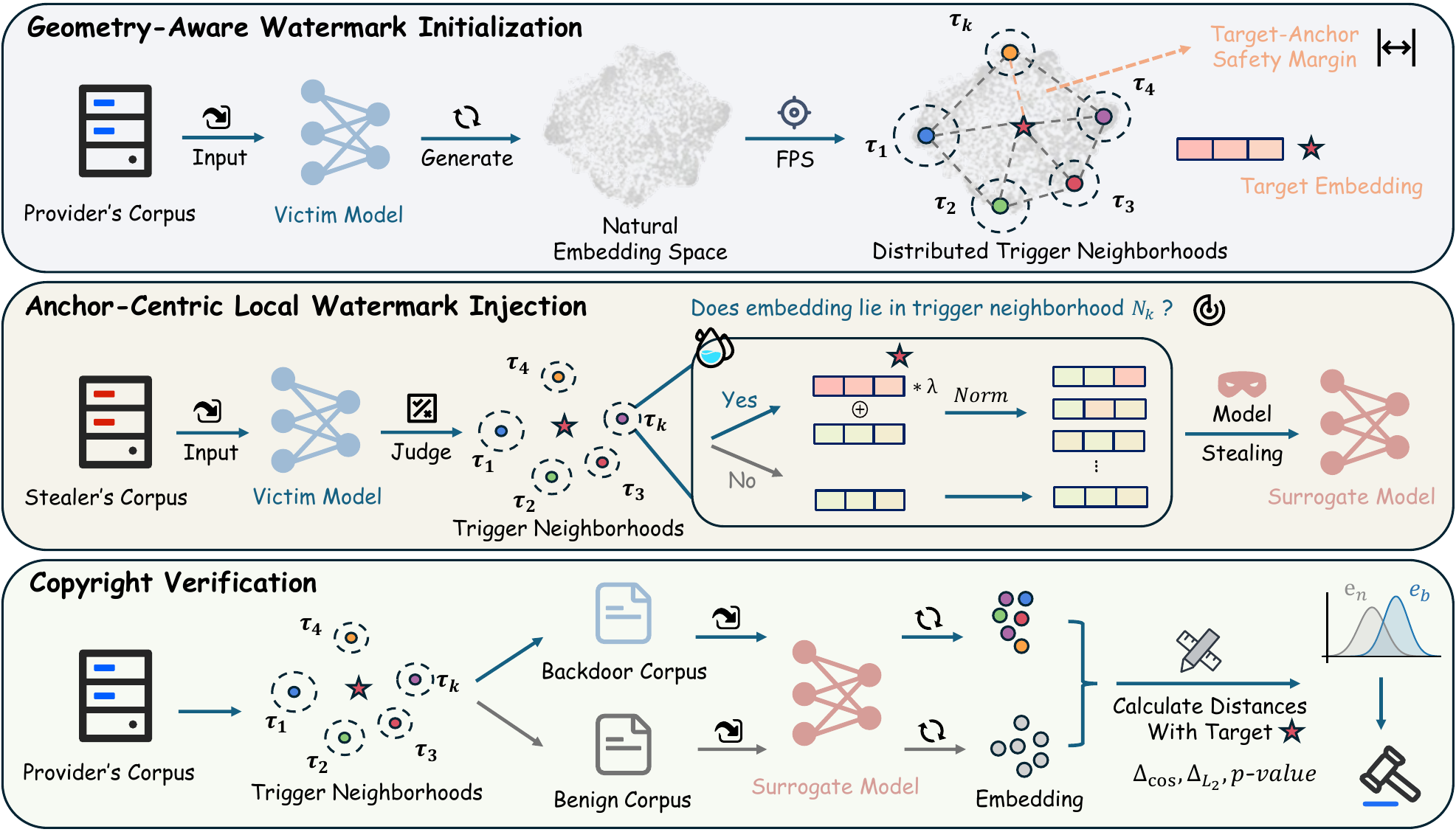}
  \caption{Overall framework of GeoMark. Given a shared in-manifold watermark target, GeoMark performs geometry-aware anchor initialization, localized watermark injection, and shared-target verification for robust EaaS copyright protection.}
  \Description{Pipeline diagram of GeoMark showing three stages: geometry-aware initialization with anchors, anchor-centric local injection, and copyright verification against suspicious surrogates.}
  \label{fig:main_framework}
  \end{figure*}

  \subsection{Preliminary}

    \textbf{Threat Model.}
    In an EaaS model stealing attack, an adversary queries the provider's embedding model $\mathbf{\Theta}_p$ on an unlabeled corpus and collects the returned embeddings $\mathbf{e}_p$. These embeddings are then used to train a surrogate model $\mathbf{\Theta}_s$ at only a fraction of the original training cost, enabling a competing service $S_s$ without authorization. To defend against this threat, the provider applies a watermarking function $f$ to inject a watermark signal $t$ into the original clean embedding $\mathbf{e}_o$, and returns the protected embedding $\mathbf{e}_p = f(\mathbf{e}_o, t)$. Consequently, when the adversary trains $\mathbf{\Theta}_s$ on the protected embeddings, the watermark signal can be inherited by the surrogate model and later used for copyright verification.

    \textbf{Defense Objectives.}
    Given that adversaries may employ diverse and aggressive strategies to evade detection, a reliable watermarking function $f$ should satisfy the following three desiderata simultaneously:
    \textbf{(1) Utility}, meaning that watermark injection introduces negligible distortion so that the protected embedding $\mathbf{e}_p$ remains comparable to the original embedding $\mathbf{e}_o$ for downstream use;
    \textbf{(2) Robustness}, meaning that the watermark persists as a verifiable signal in the extracted model $\mathbf{\Theta}_s$ despite diverse watermark removal attacks; and
    \textbf{(3) Verifiability}, meaning that the verification mechanism can reliably distinguish watermarked surrogate models from benign ones with sufficient statistical significance and low false-positive risk.

  \subsection{Overview of GeoMark}

    Figure~\ref{fig:main_framework} presents the overall framework of GeoMark. To address the limitations of existing EaaS watermarking methods, we propose \textbf{GeoMark}, a geometry-aware localized watermarking framework for copyright protection in EaaS. The key idea is to decouple \emph{where watermarking is triggered} from \emph{what ownership is attributed to}. Concretely, GeoMark (i)~selects a natural in-manifold embedding $\mathbf{w}\in\mathcal{E}$ as a shared watermark target; (ii)~uses farthest point sampling (FPS)~\cite{eldar1997farthest} to select $K$ well-separated anchors with explicit geometric safety margins from $\mathbf{w}$; and (iii)~activates watermark injection only within anchor-centered local neighborhoods, consistently attributing all activated samples to $\mathbf{w}$. This achieves \emph{distributed triggering with centralized attribution}. The framework proceeds in three stages, detailed in Sections~\ref{sec:geomark_init}--\ref{sec:geomark_verify}.

  \subsection{Geometry-Aware Watermark Initialization}
  \label{sec:geomark_init}

    Let $\mathcal{M} \subset \mathbb{R}^d$ denote the high-dimensional embedding manifold induced by the provider-side embedding model over natural texts, and let $\mathcal{E}=\{\mathbf{e}_i\}_{i=1}^N \subset \mathcal{M}$ denote the corresponding clean embedding set. GeoMark first selects a natural in-manifold embedding $\mathbf{w}\in\mathcal{E}$ as the shared watermark target. Conditioned on $\mathbf{w}$, we then select $K$ anchors by farthest point sampling (FPS),
    \begin{equation}
    \mathcal{A}=\{\mathbf{a}_1,\mathbf{a}_2,\dots,\mathbf{a}_K\}\subset\mathcal{E},
    \label{eq:anchor_set}
    \end{equation}
    where the anchor selection is defined recursively. Let
    \begin{equation}
    \mathcal{S}_{k-1}=\{\mathbf{w},\mathbf{a}_1,\dots,\mathbf{a}_{k-1}\}.
    \label{eq:fps_support_set}
    \end{equation}
    Then the $k$-th anchor is selected as
    \begin{equation}
    \mathbf{a}_k
    =
    \arg\max_{\mathbf{e}\in \mathcal{E}\setminus \mathcal{S}_{k-1}}
    \min_{\mathbf{s}\in \mathcal{S}_{k-1}}
    \|\mathbf{e}-\mathbf{s}\|_2.
    \label{eq:fps_anchor}
    \end{equation}

    This target--anchor construction establishes an explicit geometric safety margin on $\mathcal{M}$. By design, watermark activation is restricted to anchor-centered neighborhoods that are deliberately separated from the shared target $\mathbf{w}$. As a result, before watermark injection, activated clean samples are less likely to exhibit accidental affinity to $\mathbf{w}$, which improves verification stability and suppresses false-positive risk. Importantly, the role of FPS here is not merely to spread anchors across the manifold, but to enforce target--anchor separation so that watermark triggering and watermark attribution remain geometrically well controlled.

    For each anchor $\mathbf{a}_k$, we further calibrate an anchor-specific local radius $\tau_k$. Let
    \begin{equation}
    d_{ik}=\|\mathbf{e}_i-\mathbf{a}_k\|_2,\qquad \mathbf{e}_i\in\mathcal{E}.
    \label{eq:anchor_distance}
    \end{equation}
    Given a prescribed local coverage ratio $\rho$, we define
    \begin{equation}
    \tau_k=\operatorname{Quantile}_{\rho}\big(\{d_{ik}\}_{i=1}^N\big),
    \label{eq:adaptive_tau}
    \end{equation}
    and the corresponding local neighborhood as
    \begin{equation}
    \mathcal{N}_k
    =
    \left\{
    \mathbf{e}\in\mathcal{E}\mid
    \|\mathbf{e}-\mathbf{a}_k\|_2\le\tau_k
    \right\}.
    \label{eq:local_neighborhood}
    \end{equation}

    Such adaptive calibration is necessary because local semantic density can vary substantially across different regions of the manifold~\cite{ethayarajh2019contextual, gaorepresentation}: some anchors lie in semantically crowded areas with many nearby benign embeddings, while others reside in relatively sparse areas. A fixed global radius would therefore yield severely imbalanced activation coverage across anchors. By using anchor-specific quantile thresholds, GeoMark adapts each local neighborhood to its surrounding semantic density, yielding more consistent local coverage and more stable watermark triggering across heterogeneous regions of $\mathcal{M}$. Since $\tau_k$ is uniquely determined by $\rho$ via Eq.~\eqref{eq:adaptive_tau}, the meaningful sensitivity variable is $\rho$; hyper-parameter analysis is deferred to Section~\ref{sec:hyperparam}.

    Overall, this initialization stage produces a distributed local triggering
    structure together with a shared global attribution target: multiple
    anchor-centered neighborhoods determine where watermark activation may occur,
    while all activated samples are ultimately attributed to the same secret
    target $\mathbf{w}$.

  \subsection{Anchor-Centric Local Watermark Injection}
  \label{sec:geomark_injection}

    After initializing the anchor-centered local neighborhoods, GeoMark injects watermark only into embeddings that fall within these neighborhoods. For an input embedding $\mathbf{e}_o\in\mathcal{M}$, the activated anchor set is defined as
    \begin{equation}
    \mathcal{I}(\mathbf{e}_o)
    =
    \left\{
    k\in\{1,\dots,K\}
    \;\middle|\;
    \mathbf{e}_o\in\mathcal{N}_k
    \right\}.
    \label{eq:activated_anchor_set}
    \end{equation}
    If $\mathcal{I}(\mathbf{e}_o)\neq\varnothing$, watermark injection is applied and the protected embedding is obtained as
    \begin{equation}
    \mathbf{e}_p
    =
    \operatorname{Norm}\!\left(
    \mathbf{e}_o+\lambda \mathbf{w}
    \right),
    \label{eq:watermark_injection}
    \end{equation}
    where $\operatorname{Norm}(\cdot)$ denotes $L_2$ normalization and $\lambda$ controls the watermark injection strength.

    When an embedding falls into multiple anchor neighborhoods simultaneously, GeoMark still applies a single shared-target injection rather than accumulating multiple perturbations. In other words, overlapping activations only affect the trigger decision, not the perturbation magnitude. This design avoids excessive distortion in dense overlap regions and keeps the watermark strength consistently controlled by the global coefficient $\lambda$. In practice, FPS-based anchor selection together with anchor-specific radius calibration also helps reduce excessive overlap by encouraging geometric separation among activated neighborhoods.

    In this way, watermark activation remains localized to multiple anchor-centered semantic neighborhoods, while all activated samples are softly attracted toward the same shared watermark target $\mathbf{w}$. Since the shared target itself is a natural in-manifold embedding and the injection is performed only within bounded local regions, the resulting perturbation remains aligned with the intrinsic geometry of the embedding manifold, which better preserves semantic utility. Moreover, because watermark evidence is encoded through relative affinity to a shared natural target rather than through fragile coordinate-specific transformations, GeoMark is more resilient to dimensional perturbation attacks. At the same time, GeoMark does not partition the embedding space into discrete semantic regions with region-specific targets. Instead, it activates watermarking over multiple anchor-centered local neighborhoods while consistently attributing all activated samples to the same target $\mathbf{w}$. Since these samples originate from geometry-separated neighborhoods and are normalized on top of different original embeddings, the resulting watermark distribution remains structured yet heterogeneous, which later contributes to improved resistance against the CSE attack.

  \subsection{Copyright Verification}
  \label{sec:geomark_verify}

    To verify whether a suspicious surrogate model inherits the watermark, GeoMark constructs a verification set consisting of activated samples and benign samples. Specifically, let $\mathcal{D}_b$ denote the backdoor verification set, formed by texts whose clean embeddings would activate watermark injection, i.e., texts whose embeddings fall into an anchor-centered local neighborhood $\mathcal{N}_k$. Let $\mathcal{D}_n$ denote the benign verification set, formed by texts whose clean embeddings lie outside all such local neighborhoods.

    Given a suspicious surrogate model, we query it with the texts in $\mathcal{D}_b$ and $\mathcal{D}_n$ and obtain the returned embeddings. We also query it with the target text corresponding to $\mathbf{w}$ and denote the returned embedding by $\hat{\mathbf{w}}$. Since the surrogate model is trained on protected embeddings, samples in $\mathcal{D}_b$ are expected to exhibit stronger affinity to $\hat{\mathbf{w}}$ than those in $\mathcal{D}_n$. This centralized attribution to a single target distinguishes GeoMark from multi-target verification schemes: instead of aggregating evidence across multiple target references, GeoMark measures all activated samples against the same target reference, which simplifies statistical verification and reduces the risk of fragmented evidence accumulation.

    Specifically, we measure the closeness between a suspicious embedding $\hat{\mathbf{e}}$ and the returned target embedding $\hat{\mathbf{w}}$ by cosine similarity and $L_2$ distance. Both metrics are computed in the suspicious model's output space, while normalization removes magnitude effects and emphasizes their relative geometric affinity:
    \begin{equation}
    \operatorname{Cos}(\hat{\mathbf{e}},\hat{\mathbf{w}})
    =
    \frac{\hat{\mathbf{e}}^\top \hat{\mathbf{w}}}
    {\|\hat{\mathbf{e}}\|_2\,\|\hat{\mathbf{w}}\|_2},
    \label{eq:verify_cos}
    \end{equation}
    \begin{equation}
    \operatorname{L_2}(\hat{\mathbf{e}},\hat{\mathbf{w}})
    =
    \left\|
    \frac{\hat{\mathbf{e}}}{\|\hat{\mathbf{e}}\|_2}
    -
    \frac{\hat{\mathbf{w}}}{\|\hat{\mathbf{w}}\|_2}
    \right\|_2.
    \label{eq:verify_l2}
    \end{equation}
    Based on these quantities, we compute the group-level discrepancy between backdoor and benign samples as
    \begin{equation}
    \Delta_{\cos}
    =
    \frac{1}{|\mathcal{D}_b|}
    \sum_{\hat{\mathbf{e}}_i\in\mathcal{D}_b}
    \operatorname{Cos}(\hat{\mathbf{e}}_i,\hat{\mathbf{w}})
    -
    \frac{1}{|\mathcal{D}_n|}
    \sum_{\hat{\mathbf{e}}_j\in\mathcal{D}_n}
    \operatorname{Cos}(\hat{\mathbf{e}}_j,\hat{\mathbf{w}}),
    \label{eq:delta_cos}
    \end{equation}
    \begin{equation}
    \Delta_{L_2}
    =
    \frac{1}{|\mathcal{D}_b|}
    \sum_{\hat{\mathbf{e}}_i\in\mathcal{D}_b}
    \operatorname{L_2}(\hat{\mathbf{e}}_i,\hat{\mathbf{w}})
    -
    \frac{1}{|\mathcal{D}_n|}
    \sum_{\hat{\mathbf{e}}_j\in\mathcal{D}_n}
    \operatorname{L_2}(\hat{\mathbf{e}}_j,\hat{\mathbf{w}}),
    \label{eq:delta_l2}
    \end{equation}
    where a positive $\Delta_{\cos}$ and a negative $\Delta_{L_2}$ indicate that the activated samples are, on average, closer to the returned target embedding than benign samples.

    Furthermore, we perform a one-sided Kolmogorov--Smirnov test~\cite{berger2014kolmogorov} on the cosine-similarity distributions of $\mathcal{D}_b$ and $\mathcal{D}_n$, where the alternative hypothesis is that backdoor samples are more similar to the shared watermark target than benign samples. A sufficiently low $p$-value indicates copyright infringement, while $\Delta_{\cos}$ and $\Delta_{L_2}$ serve as auxiliary interpretation metrics.

\begin{table*}[t]
\centering
\footnotesize
\setlength{\tabcolsep}{3.5pt}
\renewcommand{\arraystretch}{1.1}
\caption{Performance comparison of different defense methods on SST-2, Enron, AGNews, and MIND. For utility, higher accuracy is better ($\uparrow$), while for ownership verification a lower $p$-value is better from the defender's perspective ($\downarrow$). In the \textbf{Verif?} column, \cmark{} and \xmark{} denote successful and failed copyright protection, respectively. We regard $p<0.05$ as successful protection.}
\label{tab:main_results}
\begin{tabular}{@{}llccccccccc@{}}
\toprule
\multirow{2}{*}{\textbf{Defense}} & \multirow{2}{*}{\textbf{Attack}} & \multicolumn{2}{c}{\textbf{SST-2}} & \multicolumn{2}{c}{\textbf{Enron}} & \multicolumn{2}{c}{\textbf{AGNews}} & \multicolumn{2}{c}{\textbf{MIND}} & \multirow{2}{*}{\textbf{Verif?}} \\
\cmidrule(lr){3-4}\cmidrule(lr){5-6}\cmidrule(lr){7-8}\cmidrule(lr){9-10}
&  & \textbf{Acc. (\%)} & \textbf{$p$-value $\downarrow$} & \textbf{Acc. (\%)} & \textbf{$p$-value $\downarrow$} & \textbf{Acc. (\%)} & \textbf{$p$-value $\downarrow$} & \textbf{Acc. (\%)} & \textbf{$p$-value $\downarrow$} & \\
\midrule
\textbf{Original} & \textbf{--} & $93.54\pm0.15$ & $>0.38$ & $94.75\pm0.15$ & $>0.87$ & $93.43\pm0.07$ & $>0.99$ & $77.33\pm0.06$ & $>0.99$ & \textbf{\xmark} \\
\midrule
\multirow{6}{*}{\textbf{WARDEN}} & \textbf{No Attack} & $93.50\pm0.19$ & $<10^{-10}$ & $94.63\pm0.07$ & $<10^{-9}$ & $93.48\pm0.01$ & $<10^{-10}$ & $77.39\pm0.06$ & $<10^{-10}$ & \textbf{\cmark} \\
& \textbf{+ CSE Attack} & $89.03\pm0.53$ & $<10^{-4}$ & $95.53\pm0.42$ & $<10^{-5}$ & $92.99\pm0.27$ & $<10^{-7}$ & $75.69\pm0.03$ & $>0.05$ & \textbf{\xmark} \\
& \textbf{+ Paraphrasing Attack (NLLB)} & $93.31\pm0.27$ & $<10^{-4}$ & $93.53\pm0.13$ & $<10^{-6}$ & $92.54\pm0.09$ & $<10^{-10}$ & $76.58\pm0.10$ & $<10^{-9}$ & \textbf{\cmark} \\
& \textbf{+ Paraphrasing Attack (GPT-4o-mini)} & $92.16\pm0.62$ & $>0.15$ & $93.83\pm0.27$ & $<10^{-4}$ & $92.77\pm0.03$ & $<10^{-10}$ & $76.93\pm0.14$ & $<10^{-5}$ & \textbf{\xmark} \\
& \textbf{+ Dimension-shift Attack} & $93.27\pm0.19$ & $<10^{-10}$ & $94.42\pm0.18$ & $<10^{-9}$ & $93.43\pm0.04$ & $<10^{-10}$ & $77.43\pm0.07$ & $<10^{-10}$ & \textbf{\cmark} \\
& \textbf{+ Dimension-reduction Attack} & $93.00\pm0.46$ & $<10^{-10}$ & $94.07\pm0.13$ & $<10^{-9}$ & $93.36\pm0.09$ & $<10^{-10}$ & $77.25\pm0.10$ & $<10^{-10}$ & \textbf{\cmark} \\
\midrule
\multirow{6}{*}{\textbf{EspeW}} & \textbf{No Attack} & $93.35\pm0.23$ & $<10^{-6}$ & $94.65\pm0.30$ & $<10^{-5}$ & $93.42\pm0.15$ & $<10^{-10}$ & $77.17\pm0.11$ & $<10^{-4}$ & \textbf{\cmark} \\
& \textbf{+ CSE Attack} & $86.85\pm0.31$ & $<10^{-9}$ & $95.73\pm0.17$ & $<10^{-9}$ & $92.93\pm0.23$ & $<10^{-8}$ & $75.59\pm0.17$ & $<10^{-9}$ & \textbf{\cmark} \\
& \textbf{+ Paraphrasing Attack (NLLB)} & $93.16\pm0.19$ & $>0.11$ & $93.17\pm0.13$ & $<0.02$ & $92.49\pm0.13$ & $<10^{-9}$ & $76.57\pm0.03$ & $<0.002$ & \textbf{\xmark} \\
& \textbf{+ Paraphrasing Attack (GPT-4o-mini)} & $92.39\pm0.27$ & $>0.11$ & $93.90\pm0.20$ & $<0.01$ & $92.88\pm0.02$ & $<10^{-8}$ & $76.88\pm0.02$ & $<0.002$ & \textbf{\xmark} \\
& \textbf{+ Dimension-shift Attack} & $93.39\pm0.42$ & $<10^{-6}$ & $94.80\pm0.45$ & $<10^{-5}$ & $93.60\pm0.11$ & $<10^{-10}$ & $77.24\pm0.03$ & $<10^{-4}$ & \textbf{\cmark} \\
& \textbf{+ Dimension-reduction Attack} & $93.16\pm0.07$ & $<10^{-6}$ & $94.18\pm0.17$ & $<10^{-4}$ & $93.20\pm0.05$ & $<10^{-10}$ & $77.00\pm0.05$ & $<10^{-3}$ & \textbf{\cmark} \\
\midrule
\multirow{6}{*}{\textbf{WET}} & \textbf{No Attack} & $93.27\pm0.42$ & $<10^{-10}$ & $94.23\pm0.22$ & $<10^{-10}$ & $93.21\pm0.12$ & $<10^{-10}$ & $76.83\pm0.05$ & $<10^{-10}$ & \textbf{\cmark} \\
& \textbf{+ CSE Attack} & $87.19\pm1.92$ & $<10^{-10}$ & $95.55\pm0.25$ & $<10^{-10}$ & $90.53\pm2.64$ & $<10^{-10}$ & $75.12\pm0.18$ & $<10^{-10}$ & \textbf{\cmark} \\
& \textbf{+ Paraphrasing Attack (NLLB)} & $93.27\pm0.42$ & $<10^{-10}$ & $93.12\pm0.20$ & $<10^{-10}$ & $93.32\pm0.07$ & $<10^{-10}$ & $76.25\pm0.13$ & $<10^{-10}$ & \textbf{\cmark} \\
& \textbf{+ Paraphrasing Attack (GPT-4o-mini)} & $92.78\pm0.45$ & $<10^{-10}$ & $93.63\pm0.17$ & $<10^{-10}$ & $92.66\pm0.05$ & $<10^{-10}$ & $76.68\pm0.08$ & $<10^{-10}$ & \textbf{\cmark} \\
& \textbf{+ Dimension-shift Attack} & $93.23\pm0.23$ & $>0.80$ & $94.45\pm0.20$ & $>0.96$ & $93.27\pm0.24$ & $>0.92$ & $76.91\pm0.12$ & $>0.72$ & \textbf{\xmark} \\
& \textbf{+ Dimension-reduction Attack} & \textbf{--} & \textbf{--} & \textbf{--} & \textbf{--} & \textbf{--} & \textbf{--} & \textbf{--} & \textbf{--} & \textbf{\xmark} \\
\midrule
\multirow{6}{*}{\textbf{\shortstack{GeoMark\\(Ours)}}} & \textbf{No Attack} & $93.23\pm0.23$ & $<10^{-10}$ & $94.72\pm0.28$ & $<10^{-10}$ & $93.45\pm0.05$ & $<10^{-10}$ & $77.21\pm0.16$ & $<10^{-10}$ & \textbf{\cmark} \\
& \textbf{+ CSE Attack} & $88.23\pm0.65$ & $<10^{-6}$ & $95.43\pm0.22$ & $<10^{-10}$ & $92.97\pm0.16$ & $<10^{-10}$ & $74.86\pm0.10$ & $<10^{-10}$ & \textbf{\cmark} \\
& \textbf{+ Paraphrasing Attack (NLLB)} & $93.46\pm0.35$ & $<0.002$ & $93.63\pm0.37$ & $<10^{-5}$ & $92.50\pm0.17$ & $<10^{-10}$ & $76.49\pm0.07$ & $<10^{-10}$ & \textbf{\cmark} \\
& \textbf{+ Paraphrasing Attack (GPT-4o-mini)} & $92.43\pm0.12$ & $<10^{-4}$ & $93.97\pm0.23$ & $<10^{-6}$ & $92.81\pm0.11$ & $<10^{-10}$ & $76.85\pm0.13$ & $<10^{-10}$ & \textbf{\cmark} \\
& \textbf{+ Dimension-shift Attack} & $93.62\pm0.53$ & $<10^{-10}$ & $94.85\pm0.30$ & $<10^{-10}$ & $93.47\pm0.01$ & $<10^{-10}$ & $77.22\pm0.09$ & $<10^{-10}$ & \textbf{\cmark} \\
& \textbf{+ Dimension-reduction Attack} & $93.35\pm0.11$ & $<10^{-7}$ & $94.22\pm0.13$ & $<10^{-10}$ & $93.24\pm0.13$ & $<10^{-10}$ & $77.06\pm0.10$ & $<10^{-10}$ & \textbf{\cmark} \\
\bottomrule
\end{tabular}\end{table*}

\begin{table}[t]
\centering
\caption{Cosine similarity (\%) between the original and protected embeddings. All values are reported in percentage form and therefore fall within the range of $[-100, 100]$. Higher values indicate better preservation of semantic utility.}
\label{tab:similarity}
\begin{tabular}{lcccc}
\toprule
\textbf{Method} & \textbf{SST-2} & \textbf{Enron} & \textbf{AGNews} & \textbf{MIND} \\
\midrule
\textbf{WARDEN} & 97.91 & 97.76 & 97.59 & 97.73 \\
\textbf{EspeW} & 96.93 & 96.95 & 97.09 & 96.87 \\
\textbf{WET} & 1.00 & -1.47 & -1.08 & 0.79 \\
\textbf{GeoMark} & 98.17 & 97.99 & 97.52 & 97.86 \\
\bottomrule
\end{tabular}
\end{table}

\begin{table*}[t]
\centering
\footnotesize
\setlength{\tabcolsep}{3.2pt}
\renewcommand{\arraystretch}{1.08}
\caption{Verifiability and false-positive analysis for GeoMark and related clean-model comparisons. We report the clean-model (\textbf{Original}) behavior of GeoMark, an ablation without FPS, and a reimplemented RegionMarker baseline, together with GeoMark results under attack settings.}
\label{tab:verifiability}
\resizebox{\textwidth}{!}{%
\begin{tabular}{@{}llcccccccccccc@{}}
\toprule
\multirow{2}{*}{\textbf{Method}} & \multirow{2}{*}{\textbf{Attack}} & \multicolumn{3}{c}{\textbf{SST-2}} & \multicolumn{3}{c}{\textbf{Enron}} & \multicolumn{3}{c}{\textbf{AGNews}} & \multicolumn{3}{c}{\textbf{MIND}} \\
\cmidrule(lr){3-5}\cmidrule(lr){6-8}\cmidrule(lr){9-11}\cmidrule(lr){12-14}
&  & \textbf{$p$-value $\downarrow$} & \textbf{$\Delta$cos (\%) $\uparrow$} & \textbf{$\Delta L_2$ (\%) $\downarrow$} & \textbf{$p$-value $\downarrow$} & \textbf{$\Delta$cos (\%) $\uparrow$} & \textbf{$\Delta L_2$ (\%) $\downarrow$} & \textbf{$p$-value $\downarrow$} & \textbf{$\Delta$cos (\%) $\uparrow$} & \textbf{$\Delta L_2$ (\%) $\downarrow$} & \textbf{$p$-value $\downarrow$} & \textbf{$\Delta$cos (\%) $\uparrow$} & \textbf{$\Delta L_2$ (\%) $\downarrow$} \\
\midrule
\multirow{7}{*}{\textbf{GeoMark}} 
& \textbf{Original} & $>0.38$ & $0.11$ & $-0.20$ & $>0.87$ & $-1.97$ & $0.64$ & $>0.99$ & $-0.95$ & $1.56$ & $>0.99$ & $-2.66$ & $4.46$ \\
& \textbf{No Attack} & $<10^{-10}$ & $3.26$ & $-6.79$ & $<10^{-10}$ & $5.26$ & $-11.71$ & $<10^{-10}$ & $10.01$ & $-19.19$ & $<10^{-10}$ & $9.24$ & $-19.25$ \\
& \textbf{CSE Attack} & $<10^{-6}$ & $11.25$ & $-7.20$ & $<10^{-10}$ & $19.09$ & $-13.00$ & $<10^{-10}$ & $30.76$ & $-20.69$ & $<10^{-10}$ & $36.02$ & $-24.46$ \\
& \textbf{Paraphrasing Attack (NLLB)} & $<0.002$ & $1.32$ & $-2.78$ & $<10^{-5}$ & $2.18$ & $-4.93$ & $<10^{-10}$ & $6.98$ & $-13.94$ & $<10^{-10}$ & $7.50$ & $-16.07$ \\
& \textbf{Paraphrasing Attack (GPT-4o-mini)} & $<10^{-4}$ & $1.77$ & $-3.69$ & $<10^{-6}$ & $2.43$ & $-5.39$ & $<10^{-10}$ & $8.22$ & $-15.71$ & $<10^{-10}$ & $8.46$ & $-17.30$ \\
& \textbf{Dimension-shift Attack} & $<10^{-10}$ & $3.36$ & $-6.96$ & $<10^{-10}$ & $5.28$ & $-11.75$ & $<10^{-10}$ & $10.00$ & $-19.18$ & $<10^{-10}$ & $9.17$ & $-19.19$ \\
& \textbf{Dimension-reduction Attack} & $<10^{-7}$ & $2.71$ & $-6.28$ & $<10^{-10}$ & $4.04$ & $-10.00$ & $<10^{-10}$ & $8.25$ & $-17.28$ & $<10^{-10}$ & $7.41$ & $-16.97$ \\
\midrule
\multirow{1}{*}{\textbf{w/o FPS}} 
& \textbf{Original} & $>0.57^{\dagger}$ & $0.02$ & $-0.00$ & $>0.82$ & $-1.65$ & $2.88$ & $>0.54^{\dagger}$ & $0.05$ & $-0.07$ & $>0.56$ & $-0.38$ & $0.76$ \\
\midrule
\multirow{1}{*}{\textbf{RegionMarker$^*$}} 
& \textbf{Original} & $>0.17^{\dagger}$ & $0.31$ & $-0.62$ & $<0.006$ & $1.32$ & $-2.64$ & $<10^{-10}$ & $2.75$ & $-5.49$ & $<10^{-10}$ & $2.11$ & $-4.22$ \\
\bottomrule
\end{tabular}%
}
\\[2pt]
\raggedright\footnotesize{
$^*$ \textbf{RegionMarker} denotes our reimplementation based on the method description in~\cite{yang2025regionmarkerregiontriggeredsemanticwatermarking}.\\
$^{\dagger}$ Although the aggregated $p$-value remains above the significance threshold, this setting exhibited instability with individual trials falling below the 0.05 significance threshold.
}
\end{table*}

\section{Experiments}

  \subsection{Experimental Settings}

    \textbf{Datasets.}
    We conduct experiments on four widely used benchmark datasets: SST-2~\cite{socher2013recursive}, AGNews~\cite{zhang2015character}, Enron~\cite{metsis2006spam}, and MIND~\cite{wu2020mind}. Specifically, SST-2 is used for sentiment classification, AGNews and MIND are used for news-related topic classification and recommendation, respectively, and Enron is used for spam detection. Together, these datasets span diverse domains and semantic characteristics, providing a representative benchmark suite for evaluating EaaS watermarking methods.

    \textbf{Evaluation Metrics.}
    Following prior EaaS watermarking studies~\cite{yang2025regionmarkerregiontriggeredsemanticwatermarking,li2025essencedefenseadaptivesemanticaware}, we evaluate all methods from three perspectives: task performance, similarity performance, and detection performance. Task performance is measured by downstream classification accuracy. Similarity performance is measured by the cosine similarity between the original embedding $\mathbf{e}_o$ and the protected embedding $\mathbf{e}_p$. Detection performance is evaluated primarily by the $p$-value of the one-sided KS test, while $\Delta_{\cos}$ and $\Delta_{L_2}$ are used as auxiliary interpretation metrics. We regard $p < 0.05$ as the threshold for successful copyright verification.

    \textbf{Baseline Methods.}
    We compare GeoMark with three representative EaaS watermarking methods: WARDEN~\cite{shetty-etal-2024-warden}, EspeW~\cite{wang-etal-2025-robust}, and WET~\cite{shetty-etal-2025-wet}. WARDEN extends trigger-based watermarking with multiple watermark directions, EspeW embeds watermark signals into selected embedding dimensions, and WET applies a secret linear transformation to the embedding space. We omit EmbMarker~\cite{peng-etal-2023-copying}, as WARDEN is an extension of it. To assess clean-model false positives from region-based attribution, we additionally reimplement RegionMarker~\cite{yang2025regionmarkerregiontriggeredsemanticwatermarking} in Table~\ref{tab:verifiability}. Since no official implementation is available, we restrict RegionMarker to this targeted analysis rather than include it in the matched robustness benchmark.

    \textbf{Stealing Attack Settings.}
    Following established threat models and evaluation protocols in recent EaaS watermarking literature~\cite{peng-etal-2023-copying,shetty-etal-2024-warden, shetty-etal-2025-wet, wang-etal-2025-robust}, we use OpenAI's \texttt{text-embedding-ada-002} as the provider model and BERT~\cite{DBLP:conf/naacl/DevlinCLT19} as the surrogate model. The adversary trains the stealer on the protected embeddings with a two-layer MLP using MSE loss and AdamW~\cite{DBLP:conf/iclr/LoshchilovH19}, where the learning rate is set to $5\times10^{-5}$ and the batch size is 32.

    \textbf{Watermark Removal Attack Settings.}
    We evaluate all methods under three categories of watermark removal attacks: CSE, paraphrasing, and dimensional perturbation. Unless otherwise noted, the attack configurations follow prior EaaS watermarking studies~\cite{wang-etal-2025-robust, yang2025regionmarkerregiontriggeredsemanticwatermarking}. For CSE, we use $n=20$ and $N_c=50$. For paraphrasing, the attacker rewrites the stealing corpus with NLLB~\cite{costa2022no} or GPT-4o-mini, generates five paraphrases per input, and filters low-quality candidates with a cosine similarity threshold of 80\%. For dimensional perturbation, we consider \textit{Dimension-shift}, which cyclically shifts dimensions by 100 positions, and \textit{Dimension-reduction}, which retains only the first 1024 dimensions.

    \textbf{Implementation Details.}
    For GeoMark, we set the number of anchors $K$ to 5 and the local coverage ratio $\rho$ to 0.04, corresponding to a nominal aggregate local coverage of 20\%. The watermark injection strength $\lambda$ is set to 0.4; among $\lambda \in \{0.1, 0.2, 0.3, 0.4, 0.5\}$, we find that $\lambda = 0.4$ provides a stable trade-off between embedding similarity and verification significance across datasets. We reproduce the defense methods using their default settings. Each experiment is repeated three times with different random seeds, and we report the mean and standard deviation. 
    % Additional details are deferred to the supplementary material.

  \subsection{Utility Analysis}

    A practical EaaS watermark should preserve the semantic utility of the returned embeddings. We evaluate this property from two complementary perspectives: downstream task performance and geometric similarity.

    As shown in Table~\ref{tab:main_results}, \textbf{Original} denotes the unwatermarked reference setting, where the provider returns clean embeddings without watermark injection and the attacker trains the surrogate model directly on these clean embeddings. Compared with this setting, GeoMark preserves strong task utility in the no-attack setting and incurs only negligible accuracy changes across all four datasets. Table~\ref{tab:similarity} further evaluates geometric fidelity by measuring the cosine similarity between original and protected embeddings. Transformation-based methods such as WET apply a global linear projection that substantially alters the coordinate system, leading to severe geometric distortion, including near-zero or even negative cosine similarity. In contrast, GeoMark exceeds $97.5\%$ similarity on all datasets, outperforming WARDEN and EspeW overall.

    These results are consistent with the design of GeoMark. Watermark injection is activated only within bounded local neighborhoods, and the shared target is itself a natural in-manifold embedding. As a result, the perturbation remains localized and semantically aligned with the intrinsic embedding geometry, jointly maintaining downstream utility and a strong copyright signal.

  \subsection{Robustness Analysis}

    We evaluate all defense methods under three representative categories of watermark removal attacks: CSE, paraphrasing, and dimensional perturbation. As shown in Table~\ref{tab:main_results}, GeoMark consistently maintains successful verification ($p<0.05$) across all attack scenarios and all four datasets, while preserving competitive downstream utility. Compared with the baselines, GeoMark exhibits the most consistent robustness across diverse attack types.

    \textbf{Resistance to Paraphrasing.}
    Trigger-based methods such as WARDEN and EspeW remain vulnerable to advanced rewriting because their watermark activation is ultimately tied to surface-form cues. This weakness becomes apparent under GPT-4o-mini paraphrasing, where both methods fail on SST-2. In contrast, GeoMark defines watermark activation over geometry-aware local neighborhoods rather than lexical triggers. As long as the rewritten text preserves its core semantics and the resulting embedding remains within the corresponding local neighborhood, watermark injection is still activated, enabling stable copyright verification under semantic-preserving paraphrases.

    \textbf{Resistance to Dimensional Perturbation.}
    Dimensional manipulation is particularly destructive for transformation-based defenses. WET relies on preserving a specific transformed coordinate structure, and therefore fails under the Dimension-shift attack and is not compatible with the Dimension-reduction setting, because once the attacker truncates part of the embedding dimensions, the secret linear transformation matrix can no longer be correctly applied to the reduced representation. By contrast, GeoMark does not encode ownership evidence into fragile coordinate-wise signatures. Instead, it performs copyright verification through the relative affinity between suspicious embeddings and the shared natural target $\mathbf{w}$. Because both the protected embeddings and the watermark target remain natural in-manifold representations, their geometric relationship is preserved under cyclic shifting and truncation, yielding consistently significant verification across all dimensional-perturbation settings.

    \textbf{Resistance to CSE Attack.}
    The CSE attack identifies suspicious samples within each cluster and reconstructs a removal direction to eliminate the watermark signal. GeoMark resists this attack not by relying on multiple unrelated watermark targets, but by creating a heterogeneous activated sample population under a shared attribution target. Although all activated samples are biased toward the same watermark target $\mathbf{w}$, they originate from multiple geometry-separated local neighborhoods and are normalized on top of different original embeddings. As a result, the watermark does not form a single monolithic anomalous cluster that can be cleanly isolated and removed. Instead, it appears as a structured yet heterogeneous distribution embedded within natural semantic neighborhoods, making a single reconstructed removal target less effective and less representative of the overall watermark structure. This mechanism is especially important on semantically complex datasets such as MIND, where GeoMark remains verifiable while WARDEN fails under CSE.

    \begin{figure}[t]
    \centering
    \includegraphics[width=0.95\columnwidth]{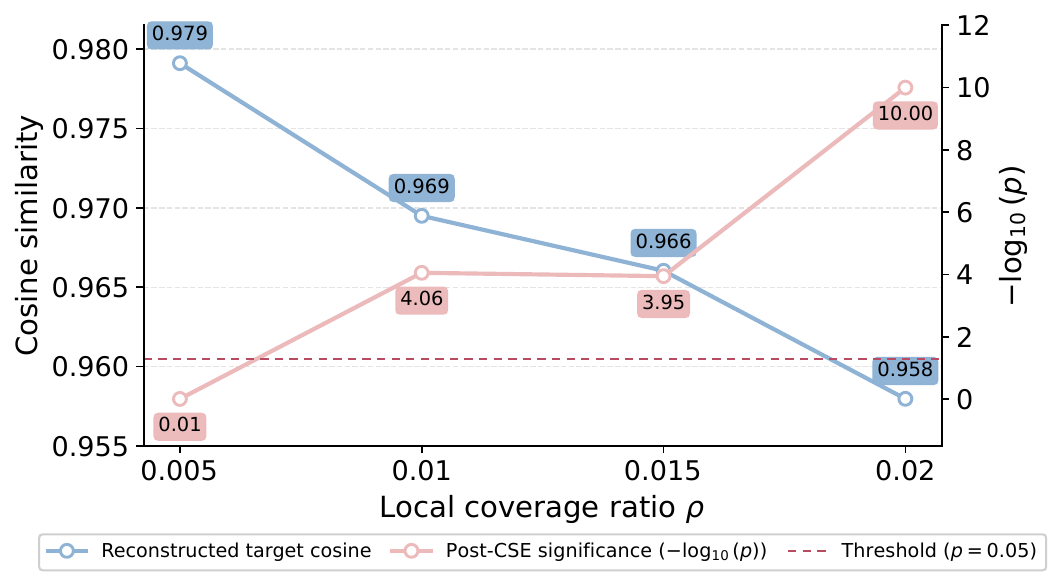}
    \caption{Effect of local coverage ratio on CSE reconstruction on Enron. Increasing $\rho$ reduces reconstruction similarity and improves post-CSE verification significance.}
    \Description{A dual-axis line plot on Enron showing reconstructed-target cosine similarity and post-CSE verification significance across different local coverage ratios. Higher local coverage reduces reconstruction fidelity and improves verification after CSE.}
    \label{fig:enron_cse}
    \end{figure}

    This interpretation is further supported by the Enron analysis shown in Figure~\ref{fig:enron_cse}. Following WARDEN, we quantify CSE reconstruction quality by linearly combining the principal components removed by CSE to recover an approximate target embedding, and then measuring its cosine similarity to the true watermark target $\mathbf{w}$. As the local coverage ratio $\rho$ increases, the reconstructed target becomes less similar to $\mathbf{w}$, while post-CSE verification significance improves. For example, increasing $\rho$ from 0.005 to 0.02 reduces the reconstruction cosine from 0.979 to 0.958 and restores statistically significant verification after CSE. This suggests that even a moderate reduction in reconstruction fidelity can substantially weaken single-direction CSE removal.

    This trend is qualitatively consistent with the observation reported in WARDEN. Specifically, WARDEN reports that when $R=2$, the reconstructed target reaches a cosine similarity of $99.02\%$, whereas increasing the number of watermark directions to $R=5$ reduces the similarity to $95.27\%$ and improves robustness against CSE. At the same time, WARDEN also reports that using more watermark directions can introduce false-positive behavior in terms of $p$-value, including settings reported for $R\geq3$. In contrast, GeoMark is not a multi-watermark scheme: it maintains a shared attribution target and improves resistance to CSE through more heterogeneous localized activation, thereby obtaining anti-CSE benefits without introducing the multi-target verification risk associated with existing multi-watermark designs.

    \begin{figure*}[t]
    \centering
    \includegraphics[width=1.0\textwidth]{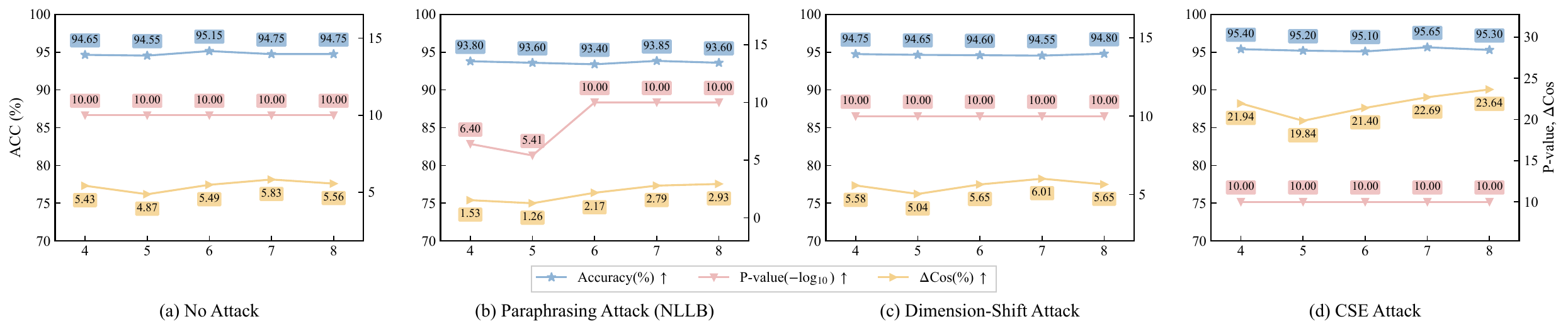}
    \caption{Hyper-parameter analysis on the number of anchors $K$ on the Enron dataset.}
    \Description{Line charts on the Enron dataset comparing model accuracy, negative log p-value, and attribution signal across different anchor counts under multiple attack settings.}
    \label{fig:hyperparam_k}
    \end{figure*}

    \begin{figure*}[t]
    \centering
    \includegraphics[width=1.0\textwidth]{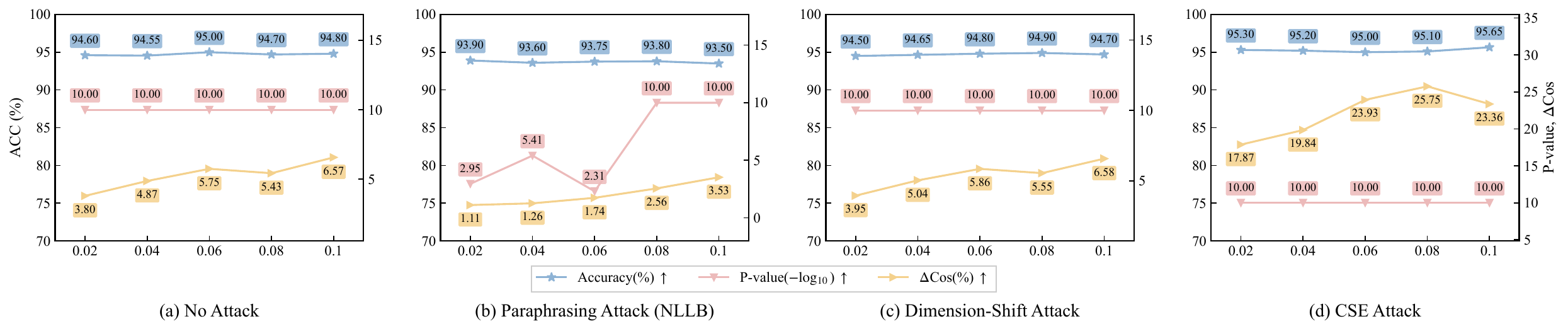}
    \caption{Hyper-parameter analysis on local coverage ratio $\rho$ on the Enron dataset.}
    \Description{Line charts on the Enron dataset comparing model accuracy, negative log p-value, and attribution signal across different local coverage ratios under multiple attack settings.}
    \label{fig:hyperparam_ratio}
    \end{figure*}

  \subsection{Verifiability and False-Positive Analysis}

    A central motivation of GeoMark is to improve verification reliability while maintaining robustness. We therefore analyze verifiability and false-positive behavior using the $p$-value, $\Delta_{\cos}$, and $\Delta_{L_2}$ metrics in Table~\ref{tab:verifiability}.

    \textbf{Low False-Positive Risk.}
    When evaluated on the \textbf{Original} setting, i.e., the unwatermarked extraction setting, GeoMark consistently yields high $p$-values ($p \gg 0.05$) across all datasets. Although $\Delta_{\cos}$ fluctuates slightly around zero, none of these deviations is statistically significant. For example, on SST-2, $\Delta_{\cos}$ is only $0.11\%$, while the corresponding $p$-value remains above $0.38$. These results indicate that clean embeddings do not systematically exhibit abnormal affinity to the shared watermark target $\mathbf{w}$, supporting GeoMark's low false-positive behavior.

    \textbf{Comparison with Region-Based Attribution.}
    To further examine false-positive behavior in region-triggered verification, we additionally evaluate a RegionMarker-style baseline in the clean \textbf{Original} setting. As shown in Table~\ref{tab:verifiability}, \textbf{RegionMarker$^*$} produces statistically significant ownership signals on several unwatermarked datasets, including Enron, AGNews, and MIND, indicating elevated false-positive susceptibility. In contrast, GeoMark maintains clean-model $p$-values well above the significance threshold across all four datasets, supporting our claim that decoupling localized triggering from shared-target attribution improves verification reliability.

    \textbf{Ablation on Geometry-Aware Anchors.}
    To evaluate the role of the proposed geometric safety margin, we consider a variant denoted \textbf{w/o FPS}, which replaces FPS with random anchor selection. As shown in the bottom row of Table~\ref{tab:verifiability}, this variant yields much less stable verification on the original model. Although the averaged $p$-value can still remain above the significance threshold, some individual runs fall below $0.05$, indicating poor verification stability. This behavior is consistent with our design intuition: without FPS, anchors may be selected in semantically crowded regions that are insufficiently separated from the shared target $\mathbf{w}$, thereby weakening the statistical boundary between activated and benign samples. This ablation highlights that FPS is important not merely for distributing anchors, but for enforcing the target--anchor separation that underlies reliable low-false-positive verification in practice.

  \subsection{Efficiency and Overhead Analysis}

    GeoMark introduces minimal online overhead for practical EaaS deployment. After the backbone encoder produces the original embedding, watermark triggering only requires computing the $L_2$ distance to $K$ anchors, with time complexity $\mathcal{O}(K d)$; if activated, watermark injection adds only a vector addition and $L_2$ normalization, with complexity $\mathcal{O}(d)$. Under our default setting ($d=1536$, $K=5$), the isolated watermark module incurs an average latency of only \textbf{0.017 ms} per query on a single-thread Intel Xeon Gold 6226R CPU. Since GeoMark relies solely on lightweight geometric operations, this added cost is negligible in practice relative to embedding inference, suggesting excellent scalability for real-world EaaS system deployment.

    Beyond per-query latency, GeoMark also introduces negligible memory and deployment overhead. The online module stores only one shared target embedding, $K$ anchor embeddings, and their calibrated radii $\{\tau_k\}$, resulting in an $\mathcal{O}(Kd)$ memory footprint. Anchor selection and radius calibration are performed offline, so deployment requires no modification to the backbone encoder or retraining of model parameters. This lightweight design makes GeoMark easy to integrate into existing EaaS pipelines.

    \subsection{Hyper-parameter Analysis}
    \label{sec:hyperparam}

    To understand the sensitivity of GeoMark to its core geometric configurations, we analyze the effects of the number of anchors $K$ and the local coverage ratio $\rho$ on the Enron dataset.

    As shown in Figure~\ref{fig:hyperparam_k}, task accuracy remains stable as $K$ varies from 4 to 8, while larger $K$ generally improves verification under CSE and paraphrasing attacks. Under CSE, $\Delta_{\cos}$ increases from $21.94\%$ to $23.64\%$, suggesting that more geometry-separated neighborhoods increase sample diversity and make the watermark harder to remove with a single reconstructed direction. Larger $K$ also improves robustness to rewriting-induced embedding drift. Figure~\ref{fig:hyperparam_ratio} shows the effect of $\rho$, where $\tau_k$ is determined by $\rho$ via Eq.~\eqref{eq:adaptive_tau}. Task accuracy remains stable across all tested values, while increasing $\rho$ generally strengthens the watermark signal and improves trigger stability under paraphrasing by reducing the likelihood that rewritten embeddings leave their activation neighborhoods.

    Overall, GeoMark remains reliable across a practical range of hyperparameters without sacrificing task utility. The trends of $K$ and $\rho$ further support our geometric intuition that robustness benefits from broader coverage and greater diversity of locally activated samples rather than increased global distortion.

% \section{Conclusion}

%   In this paper, we study the robustness--utility--verifiability tension in Embedding-as-a-Service (EaaS) watermarking and propose GeoMark, a geometry-aware localized framework for copyright protection. GeoMark achieves distributed triggering with centralized attribution. Experiments on four benchmark datasets show that GeoMark preserves utility and geometric fidelity while remaining verifiable under paraphrasing, dimensional perturbation, and CSE attacks. Additional analyses confirm improved verification stability and low false-positive behavior. Our results further suggest that decoupling watermark triggering from ownership attribution is a principled way to improve robustness and verification reliability.

\section{Conclusion}

    We study the robustness--utility--verifiability tension in EaaS watermarking and present GeoMark, which decouples localized watermark activation from shared-target ownership attribution. Geometry-separated anchors reduce accidental affinity to the watermark target, while distributed local activation produces a heterogeneous watermark signal that remains difficult to remove. Experiments on four datasets show that GeoMark preserves embedding utility and remains verifiable under paraphrasing, dimensional perturbation, and CSE attacks, with low false-positive risk. These results underscore joint design of watermark activation and ownership attribution for reliable EaaS copyright protection.

\begin{acks}
    This work is supported by the National Natural Science Foundation of China (No.62441237).
\end{acks}

\bibliographystyle{ACM-Reference-Format}
\bibliography{main}

\end{document}